\documentclass[a4paper,aps,prd,10pt,preprintnumbers,showpacs,twocolumn,superscriptaddress,nofootinbib,amsmath,amssymb]{revtex4-1}
\usepackage{graphicx}
\usepackage[utf8]{inputenc}
\usepackage[T1]{fontenc}
\usepackage{cmap}
\usepackage{ulem}

\begin{document}

\title{Telling late-time tails for a massive scalar field in the background of brane-localized black holes}
\author{Alexey Dubinsky}
\email{alexeydubinsky@us.es}
\affiliation{University of Seville, 41009 Seville, Spain}

\begin{abstract}
We examine perturbations of a massive scalar field around spherically symmetric, brane-localized black holes. Although the ringdown and asymptotic tails of various brane-world black holes have been extensively studied, there has been no analysis of the massive late-time tails for the simplest Schwarzschild-like, brane-localized black hole to date. We demonstrate that after the ringdown phase, two stages of oscillatory tails emerge – intermediate and asymptotic. The asymptotic decay law is distinct from those associated with Schwarzschild or Reissner-Nordstr\"{o}m solutions. Specifically, during intermediate times, the signal decays as $\sim t^{-(3/2) - \ell} \sin (F(\mu) t)$, while the asymptotic decay law is $\sim t^{-1} \sin (F(\mu) t)$.
\end{abstract}


\maketitle

\section{Introduction}

The perturbation of matter and gravitational fields around black holes constitutes an active area of research, propelled by recent advancements in observing various radiation phenomena in both gravitational \cite{LIGOScientific:2016aoc, LIGOScientific:2017vwq, LIGOScientific:2020zkf, Babak:2017tow} and electromagnetic \cite{EventHorizonTelescope:2019dse, Goddi:2016qax} spectra. The evolution of black hole perturbations can be conditionally divided into three stages: the initial outburst, contingent on the specifics of the perturbation process; the period of damped oscillations; and the late-time tails. The latter two stages are contingent on the black hole parameters, such as mass, angular momentum, or charge, thereby facilitating the distinction between various black hole geometries. The tail stage exhibits a stark contrast between massless and massive fields. Massless fields decay following power-law behavior, whereas massive fields display oscillatory tails \cite{Konoplya:2011qq}. The oscillatory nature of massive tails renders them, at least in principle, plausible candidates for future observations \cite{NANOGrav:2023hvm, Konoplya:2023fmh}.

Perturbations and quasinormal modes of massive fields with various spins have undergone extensive study (see, for example, \cite{Konoplya:2004wg, Konoplya:2006br, Aragon:2020teq, Ponglertsakul:2020ufm, Konoplya:2013rxa, Ohashi:2004wr, Zhang:2018jgj, Konoplya:2018qov, Gonzalez:2022upu, Ponglertsakul:2020ufm, Konoplya:2017tvu, Zhidenko:2006rs, Konoplya:2005hr, Burikham:2017gdm} and references therein). This is due to several intriguing aspects and interesting features of the spectrum when the massive term is tuned. Firstly, the effective mass appears in the perturbation equations of certain higher-dimensional scenarios, influenced by the bulk on the brane \cite{Seahra:2004fg}. The existence of massive fields allows for arbitrarily long-lived frequencies at specific values of the field's mass \cite{Ohashi:2004wr}, \cite{Konoplya:2004wg}. This phenomenon is quite broad, encompassing various spins of the field \cite{Konoplya:2005hr, Fernandes:2021qvr, Konoplya:2017tvu, Percival:2020skc}, black hole backgrounds \cite{Konoplya:2006br, Konoplya:2013rxa, Zhidenko:2006rs, Zinhailo:2018ska, Konoplya:2019hml, Bolokhov:2023bwm}, and even other types of compact object such as wormholes \cite{Churilova:2019qph}. Lastly, perturbing a massless field in the vicinity of a black hole immersed in a magnetic field leads to the emergence of an effective mass term \cite{Konoplya:2007yy, Konoplya:2008hj}, \cite{Wu:2015fwa}.

The late-time tails of massive (or effectively massive) fields have undergone extensive examination in \cite{Jing:2004zb, Koyama:2001qw, Moderski:2001tk, Konoplya:2006gq, Rogatko:2007zz, Koyama:2001ee, Koyama:2000hj, Churilova:2019qph, Gibbons:2008rs} for various black hole and wormhole solutions, as well as for some models of higher-dimensional gravity \cite{Ishihara:2008re, Gibbons:2008gg}. In \cite{Zhidenko:2006rs}, quasinormal modes of a massive scalar field in the background of the D-dimensional Schwarzschild black hole were studied, while \cite{Abdalla:2006qj, Kanti:2006ua, Kanti:2005xa, Zhidenko:2008fp} considered the quasinormal modes of matter fields on our $(3+1)$-dimensional brane. However, to the best of our knowledge, no analysis of late-time behavior for massive fields has been conducted thus far for brane-localized black holes. In this letter, we fill this gap and demonstrate that the late-time tails around brane-localized black holes follow different laws compared to those for the $(3+1)$-dimensional Schwarzschild solutions.

This letter is organized as follows. In Sec. II we consider the black hole metric and the wave-like equation, while Sec. III is devoted to constructing the time-domain profiles and fitting them to find the laws of the late-time evolution of a massive scalar field. We briefly summarized the obtained results in Sec. III. 

\section{The black hole metric and the wave equation}
The metric of the dilaton black hole is given by the following line element,
\begin{equation}\label{metric}
ds^2=-f(r)dt^2+\frac{dr^2}{f(r)}+r^2 (d\theta^2+\sin^2\theta d\phi^2),
\end{equation}
where  the metric function of the $D$-dimensional black hole projected onto the $3+1$-dimensional brane \cite{Kanti:2004nr,Kanti:2006ua} is 
\begin{equation}
f(r) = 1- \left(\frac{r_{0}}{r}\right)^{D-3} = 1- \frac{M}{r^{D-3}}.
\end{equation}
Here $M$ is the mass parameter. The matter fields are supposed to be propagating along the $(3+1)$-dimensional brane, while the gravitational field is distributed among all $D$-dimensions. 
The massive Klein-Gordon equation in a curved spacetime
\begin{equation}
\frac{1}{\sqrt{-g}}\partial_\mu \left(\sqrt{-g}g^{\mu \nu}\partial_\nu\Phi\right) - \mu^2 \Phi =0,
 \end{equation}
 can be reduced to the master wave-like equation
\begin{equation}\label{wave-equation}
\dfrac{\partial^2 \Psi}{\partial r_*^2} - \frac{\partial^2 \Psi}{\partial t^2} -V(r) \Psi=0,
\end{equation}
after the separation of variables and changing of the wave function. Here the ``tortoise coordinate'' $r_*$ is defined as:
\begin{equation}
dr_*\equiv\frac{dr}{f(r)}.
\end{equation}
The effective potential has the following form:
\begin{equation}
V(r) = f(r) \frac{\ell(\ell+1)}{r^2}+\frac{f(r)}{r}\frac{d f(r)}{dr} + f(r) \mu^2,
\end{equation}
where $\ell$ is the multipole number, which arises from the separation of the angular variables $\theta$ and $\phi$, and $\mu$ is the mass of the field.
\begin{figure}
\resizebox{\linewidth}{!}{\includegraphics{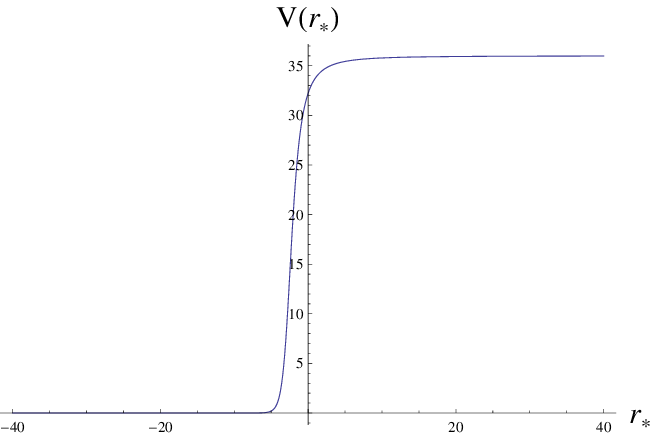}}
\caption{A typical plot of the effective potential as a function of the tortoise coordinate: $r_{0}=1$, $r_{0} \mu =6$, $\ell=1$.}\label{fig1}
\end{figure}
\begin{figure*}
\resizebox{\linewidth}{!}{\includegraphics{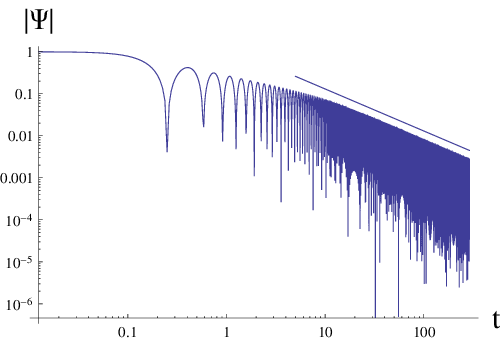}\includegraphics{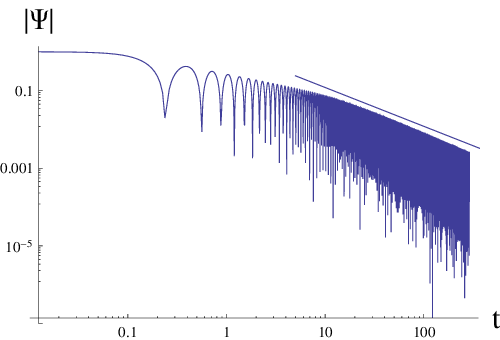}}
\caption{Logarithmic plots of the time domain profiles and the straight line parallel to the corresponding power-law envelope. Left: $\mu =10$, $\ell=1$, $D=5$, $|\Psi| \sim t^{-1}$. Right: $\mu =10$, $\ell=1$, $D=6$, $|\Psi| \sim t^{-1}$.}\label{fig2}
\end{figure*} 
\begin{figure*}
\resizebox{\linewidth}{!}{\includegraphics{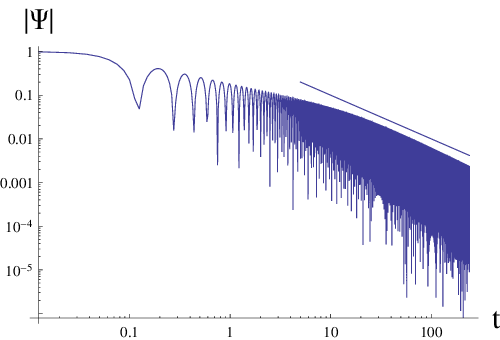}\includegraphics{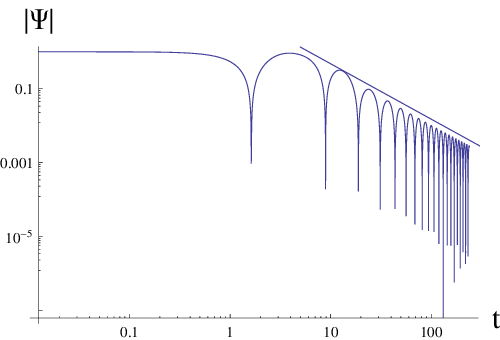}}
\caption{Logarithmic plots of the time domain profiles and the straight line parallel to the corresponding power-law envelope. Left: $\mu =20$, $\ell=0$, $D=7$, $|\Psi| \sim t^{-1}$. Right: $\mu =0.25$, $\ell=0$, $D=5$, $|\Psi| \sim t^{-3/2}$.}\label{fig3}
\end{figure*}

\begin{figure*}
\resizebox{\linewidth}{!}{\includegraphics{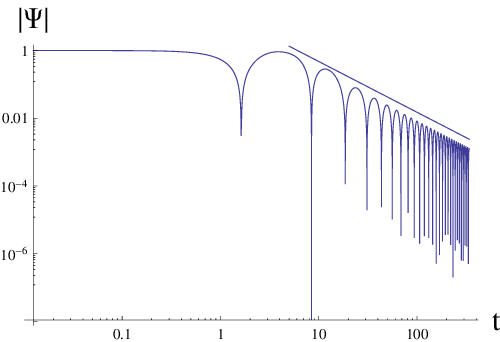}\includegraphics{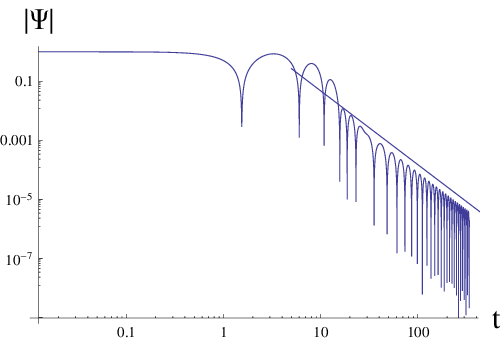}}
\caption{Logarithmic plots of the time domain profiles and the straight line parallel to the corresponding power-law envelope. Left: $\mu =0.25$, $\ell=0$, $D=6$, $|\Psi| \sim t^{-3/2}$. Right: $\mu =0.25$, $\ell=1$, $D=5$, $|\Psi| \sim t^{-5/2}$.}\label{fig4}
\end{figure*}

\section{Time-domain integration and tails}

The evolution of perturbations for asymptotically flat black holes (and some other compact objects, such as wormholes \cite{Bronnikov:2021liv}) obeys the boundary condition, requiring the purely ingoing wave at the black hole event horizon and purely outgoing waves at infinity.
Asymptotic tails could be constructed for this process via integration of the above wavelike equation in the time domain for a specific value of the radial coordinate. For this purpose we use the null-cone variables $u=t-r_*$ and $v=t+r_*$ and  the Gundlach-Price-Pullin discretization scheme \cite{Gundlach:1993tp}:
$$\Psi\left(N\right)=\Psi\left(W\right)+\Psi\left(E\right)-\Psi\left(S\right) $$
\begin{equation}\label{Discretization}
- \Delta^2V\left(S\right)\frac{\Psi\left(W\right)+\Psi\left(E\right)}{4}+{\cal O}\left(\Delta^4\right).
\end{equation}
Here, the points are designated in the following way: $N\equiv\left(u+\Delta,v+\Delta\right)$, $W\equiv\left(u+\Delta,v\right)$, $E\equiv\left(u,v+\Delta\right)$, and $S\equiv\left(u,v\right)$. We imply that some Gaussian initial data are imposed on the two null surfaces $u=u_0$ and $v=v_0$. 
This method was used in a great number of works (see, for example, \cite{Konoplya:2005et, Molina:2003dc, Konoplya:2014lha, Bolokhov:2023dxq, Churilova:2021tgn}) and proved good convergence and accuracy. Therefore, instead of repeating the well-known deductions, we refer the reader to those works and references therein.

At asymptotically late times, massless scalar and gravitational fields decay according to the Price power-law  \cite{Price:1972pw}. These massless tails do not oscillate and depend on the multipole number $\ell$.  On the contrary, for a massive scalar field either in Schwarzschild or Reissner-Nordstrom backgrounds, the asymptotic tail is oscillatory and does not depend on $\ell$ \cite{Koyama:2000hj, Koyama:2001qw}:
\begin{equation}
|\Psi| \sim t^{-5/6} \sin (\mu t), \quad t \rightarrow \infty.
\end{equation}
The regime of asymptotic times occurs at 
\begin{equation}
\frac{t}{M} > (\mu M)^{-3}.
\end{equation}
This law was observed not only for the scalar field in the Scharzschild/Reissner-Nordstr\"{o}m background, but also for a wide range of other backgrounds and fields: for a massive scalar field in the dilatonic black hole background \cite{Moderski:2001tk}, for a massive vector field in the Schwarzschild spacetime \cite{Konoplya:2006gq}, for the massive Dirac field \cite{Jing:2004zb}, for gravitational field in the Randall-Sundrum-type models \cite{Seahra:2004fg}, etc.. Nevertheless, from Figs. (\ref{fig1}-\ref{fig3}), built here via time-domain integration, we see that the asymptotic decay law is different for the brane-localized black hole:
\begin{equation}
|\Psi| \sim t^{-1} \sin (F(\mu) t), \quad t \rightarrow \infty.
\end{equation}
Here $F(\mu)$ is a function, which could be obtained approximately when fitting the numerical data for time-domain profiles in some range of values $\mu$.  
Such a distinctive behavior must be connected to a different asymptotic behavior of the black hole geometry under consideration. An analytical proof of these laws, similar to the approach using Green functions in \cite{Ching:1995tj}, might be achievable.

At the intermediate times, the tails could be observed if one chooses a relatively small value of $\mu M$ (see Figs. (\ref{fig3}-\ref{fig4})). The decay law is
\begin{equation}
|\Psi| \sim t^{-(\frac{3}{2}+\ell)} \sin (F(\mu) t).
\end{equation}
This decay law coincides with the one for the Schwarzschild case found in \cite{Hod:1998ra}.

\section{Discussions}

We have demonstrated that following the ringdown phase, the oscillatory late-time tails of a massive scalar field propagating on a $(3+1)$-dimensional brane in the background of Schwarzschild-like, brane-localized black holes are independent of the number of spatial dimensions $D$. Two types of tails exist – intermediate and asymptotic. The latter also exhibits independence from the multipole moment $\ell$. As evident from the provided time-domain profiles, the tails exhibit relatively slow decay and strong oscillation, signifying a substantial amount of energy emission during the tail stage, extending beyond the ringdown phase. The decay law at asymptotically late times is different from the well-known laws for the Schwarzschild or Reissner–Nordstr\"{o}m solutions.

\acknowledgments
The authors are thankful to R. A. Konoplya for his significant contribution to this research.

\bibliographystyle{unsrt}
\bibliography{Bibliography}

\end{document}